\documentclass[12pt]{article}
\usepackage{amssymb}
\usepackage{amsfonts}
\usepackage{amsmath,amsthm}
\usepackage{graphics,epsfig, color}
\usepackage{subfigure}

\oddsidemargin 10mm \numberwithin{equation}{section}
\def\be{\begin{equation}}
\def\ee{\end{equation}}
\begin{document}
\begin{center} {{\bf {Hairy black holes and holographic heat engine}}\\
 \vskip 0.50 cm
  {{ H. Ghaffarnejad {${}^{\dag}$} \footnote{E-mail: hghafarnejad@semnan.ac.ir
 } }{ E. Yaraie {${}^{\dag}$}\footnote{E-mail: eyaraie@semnan.ac.ir
 }}}{{  M. Farsam {${}^{\dag}$}\footnote{E-mail: mhdfarsam@semnan.ac.ir
 } }{K. Bamba{${}^{\ddag}$} \footnote{E-mail:
bamba@sss.fukushima-u.ac.jp
 }
 }}
  \vskip 0.2 cm \textit{${}^{\dag}$\textit{Faculty of Physics, Semnan
University, P.C. 35131-19111, Semnan, Iran }}
 \vskip 0.2 cm \textit{${}^{\ddag}$ Division of Human Support System, Faculty of Symbiotic Systems Science,
Fukushima University, Fukushima 960-1296, Japan }}
\end{center}
\begin{abstract}
By considering AdS charged black hole in the context of extended
thermodynamic as the working substance we use it as a heat engine.
We investigate the effect of hairy charge on the evolution of
efficiency and Carnot efficiency along with electric charge.
Because of interesting thermodynamic
 behavior of hairy black holes it would be natural to know their effects when we use black hole as a heat engine.
  We show that the hairy charge  increases the efficiency, and so maximum
temperature would be happened for bigger Maxwell charge when this
hairy charge grows.
  For the fixed electric charges, the efficiency has a minimum value. In fact all critical points describe physical states except when the charge removed.
  If the electric charge takes a zero value then the hairy charge must be negative.
   We also seek behavior of the system for large charges which is provided
 a model with low-temperature thermodynamics.
\end{abstract}
\section{Introduction}

Quantum field theory (QFT) propagated in curved spacetimes  was
studied from decade of 1970 where there is found a relationship
between geometrical and thermodynamical features of black hole,
for instance one can obtain a relation between the event horizon
surface and its entropy. In fact it is  a relation between the
surface gravity and the temperature [1,2]. This implicated a basic
connection among quantum theories, thermodynamics and gravity that
led to the birth of black hole thermodynamics laws. Hawking and
Page showed in 1983, that there is a phase transition between a
thermal AdS spacetime and Schwarzschild black hole [3]. Chamblin
\textit{et al} tested in 1999, it for a charged AdS background in
a canonical ensemble and obtained a phase transition between small
and large black holes which is similar to the Van der Waals (VdW)
phase transition [4,5]. Despite such results,  due to the lack of
a specific definition for the pressure
 and the volume of black holes this similarity is not perfect, while VdW fluid presents a first order kind of the phase transition in $P-V$ diagram.\\
In the other side Smarr relation [6,7] is not compatible with the
black hole thermodynamics first law. After some pioneer works to
promote the cosmological constant as a thermodynamic variable [8],
Kastor in 2009 tried to resolve this problem by considering a
varyiable cosmological parameter as pressure in the black hole
thermodynamics first law [9]. He reached to generalized version of
the first law of the black hole thermodynamics by considering the
cosmological parameter as origin of the thermodynamic pressure
which is conjugated quantity for the space time volume.
 By attention to these considerations
 one can see various behaviors and characteristics for black hole solutions similar to thermodynamic
 systems such as phase behaviors in gels and polymers
 or triple point in water. One of these aspects is behavior of heat engine which can be explored in black hole object.\\
From AdS/CFT duality, a gravity theory defined in AdS bulk
spacetime corresponds to a conformal field theory defined on the
boundary of AdS bulk spacetime with one dimension less.
  The cosmological constant will be related to a length scale $\ell$ in
  the d-dimensional bulk spacetime through $\Lambda=-(d-1)(d-2)/2\ell^2$.
  This length scale could be
set by the number of coincident branes in which AdS/CFT
correspondence becomes significant for its large numbers where the
curvature of spacetime becomes small. In the field theory the
number, $N$, determines degrees of freedom of boundary theory. So
dynamical pressure in the bulk leads to dynamical $N$ on the
boundary corresponds to the changing $IR$ fixed points of theory
or triggering a $RG$ flow on the boundary theory [10,11]. We can
see from [12] when the black hole is considered as a working
substance this relationship between $P$ and $N$ causes a
correspondence between a heat engine in AdS and a cycle on the
space of dual field theory side. Indeed the efficiency of heat
engine might characterize a physical property for instance a near
equilibrium expansion of an appropriate response function on the
boundary [13]. In the case of hyperbolic black holes given by
[14], heat flows and also work in field theory side are described
as the difference in the entanglement entropy of two boundary
theories which visit each other along the cycle. It must be also
interesting to study holographic heat engine at critical point at
which its efficiency is very similar to Carnot engine's at finite
power [15].
We can find a wide range of works in [14-30] at which the black hole is considered as a working substance. \\
Pioneering works of Johnson for charged AdS black holes  indicate
that mechanical work can be extracted from heat energy via "$PdV$"
term in the first law of thermodynamics.
 This is in contrary with Penrose process which leads to the extracting of energy from rotating black
 holes in both asymptotically AdS and flat spacetimes.\\
 Of course it must be noted that
some authors [31-33] found  black hole metrics with conformal
scalar hair in 4 dimensions in the case of vanishing cosmological
constant but these scalar field configurations suffer from
divergence at the horizon. There are not any black hole solutions
for higher dimensions in this case [34,35]. Applying a
non-vanishing cosmological constant, there are obtained some
conformal hairy regular black hole solutions in 3 and 4 dimensions
[36-40], but not for higher dimensions [41]. It is proved [42-44]
that if the scalar field gets coupled conformally invariant to the
higher order Euler densities then hairy black hole solution does
exist for any dimension. Motivation of considering scalar fields
has always been expected in theoretical physics and has played a
fundamental role in the structure of string theory. Since the pure
gravity is inconsistent at quantum mechanical level, so any theory
comes to replace it must be included such additional scalar field
due to its fundamental structure. In the other side scalar fields
could be useful to understand black hole physics specially in
higher dimensional black hole as a low energy effective string
theory.
 Thermodynamics of the hairy black hole is studied in [45] at which phase transition
  including backreaction is solved explicitly.
We consider the scalar hair field in a 5 dimensional RN black hole
and study its holographic heat engine when the black hole behaves
as a working substance.
 As a future work, we are interested to study thermodynamics of hairy black holes obtained from other alternative gravity models such as [46-52].\\
 Layout of this work is as follows.
In section 2 we set up hairy black hole solution and its
thermodynamic aspects in $AdS_5$ space. We discuss relationship
between the hairy and electric charges and seek the conditions for
critical points. In section 3 we calculate the efficiency of the
heat engine by regarding the effects of hairy and Maxwell charge.
We find various areas of efficiency which is restricted due to the
sign of Hawking temperature and the black hole entropy. We also
study Carnot efficiency and its evolution with respect to the
efficiency of heat engine. Finally, we can show that both of these
efficiencies approach to each other for large limit of hairy
charge. At last section we present conclusion of this work and
suggest some outlooks.

\section{Thermodynamics of hairy black holes in  a revisited $AdS_5$ space}

~~ To study hairy black holes in $AdS_5$ as a thermodynamic
 system we begin with the action as follows [44]:
\begin{equation}
I=\frac{1}{\kappa}\int d^{5}x\sqrt{-g}\bigg(
R-2\Lambda-\frac{1}{4}\mathcal{F}^2
+\kappa\mathcal{L}\left(  \phi,\nabla\phi\right)  \bigg)  \label{I}%
\end{equation}
where $\kappa=16\pi G$ is the Newton`s gravity coupling constant,
$\mathcal{F}^2=F_{\mu\nu}F^{\mu\nu}$ is the Lagrangian density of
the electromagnetic fields
 and $\mathcal{L}\left(
\phi,\nabla\phi\right)$ represents the Lagrangian of a conformally
coupled real scalar field. In general, to express the conformal
matter content, it is convenient to introduce the four-rank tensor
\begin{align} S_{\mu\nu}^{\quad\gamma\delta} &
=\phi^{2}R_{\mu\nu}^{\quad\gamma\delta
}-12\delta_{\lbrack\mu}^{[\gamma}\delta_{\nu]}^{\delta]}\nabla_{\rho}%
\phi\nabla^{\rho}\phi-\nonumber\\
&  48\phi\delta_{\lbrack\mu}^{[\gamma}\nabla_{\nu]}\nabla^{\delta]}%
\phi+18\delta_{\lbrack\mu}^{[\gamma}\nabla_{\nu]}\phi\nabla^{\delta]}%
\phi\label{Sij}%
.\end{align} In the latter case the Lagrangian density of a
conformally coupled real scalar field in 5D  is expressed by [44]
\begin{equation}
\mathcal{L}(\phi,\nabla\phi)=\phi^{15}\left(  b_{0}\mathcal{S}^{(0)}+b_{1}%
\phi^{-8}\mathcal{S}^{(1)}+b_{2}\phi^{-16}\mathcal{S}^{(2)}\right)  \label{np}%
\end{equation}
where
\begin{align*}
\mathcal{S}^{(0)}  &  =1,\\
\mathcal{S}^{(1)}  &  =S\equiv g^{\mu\nu}S_{\mu\nu}= g^{\mu\nu}%
\delta_{\sigma}^{\rho}S_{\ \mu\rho\nu}^{\sigma},\\
\mathcal{S}^{(2)}  &
=S_{\mu\nu\alpha\beta}S^{\mu\nu\alpha\beta}-4S_{\mu\nu
}S^{\mu\nu}+S^{2},
\end{align*}
 which contain higher order curvature
couplings.  In fact such couplings cause  why this model happens
to circumvent no-hair theorems [42]. We should notice that all
terms  of the Lagrangian
 (2.3) are well-behaved in the
limit $\phi\to0$  since the tensor field (2.2) is quadratic  with
respect to the scalar field [42].
 $b_0$, $b_1$ and $b_2$
 given in (2.2) are coupling constants.
They are conformally invariant under the following Weyl rescaling
transformations.
\begin{equation}
g_{\mu\nu}\rightarrow\Omega^{2}g_{\mu\nu},\qquad\phi\rightarrow\Omega
^{-1/3}\phi.
\end{equation}
Applying (2.4), the tensor field (\ref{Sij}) reads [44]
\begin{equation}
S_{\mu\nu}^{\quad\gamma\delta}\rightarrow\Omega^{-8/3}S_{\mu\nu}^{\quad
\gamma\delta}.
\end{equation}
Varying the action (\ref{I}) with respect to the metric field
$g_{\mu\nu}$ we can obtain the
 metric field equations
\begin{equation}
R_{\mu\nu}-\frac{1}{2}Rg_{\mu\nu}+\Lambda g_{\mu\nu}=\kappa
T_{\mu\nu,
}\ \label{GT}%
\end{equation}
in which the energy momentum tensor is given by
\begin{equation}
T_{\mu}^{\nu}   =\sum_{k=0}^{2 }\bigg(\frac{k!b_{k}%
}{2^{k+1}}\phi^{15-8k}\delta_{\lbrack\mu}^{\nu}\delta_{\rho_{1}}^{\lambda_{1}%
}...\delta_{\rho_{2k}]}^{\lambda_{2k}}\bigg)\times\bigg(\ S_{\ \ \ \ \lambda_{1}\lambda_{2}}^{\rho_{1}\rho_{2}%
}...S_{\ \ \ \ \lambda_{2k-1}\lambda_{2k}}^{\rho_{2k-1}\rho_{2k}}\
\bigg).
\label{TTT}
\end{equation}
Giribet et al  obtained  a spherically symmetric static hairy
black hole solution for (2.6) in 5 dimension as follows (see [44]
and references therein).
\begin{equation}
ds^{2}=-f(r)\ dt^{2}+\frac{dr^{2}}{f(r)}+r^{2}d\Omega_{3}^{2}
\label{g1}%
\end{equation}
with
\begin{equation}
f(r)=1-\frac{m}{r^{2}}-\frac{h}{r^{3}}+\frac{q^2}{r^4}-\frac{\Lambda}{6}r^{2},
\end{equation}
in which $d\Omega_{3}^{2}$ is the line element of the unit
\thinspace$3$-sphere  for which $\omega_3=\int
d\Omega_{3}^{2}=2\pi^{2}$.
 Here $m$ is mass parameter of the system and $q$ is
 electric charge  parameter  with  Maxwell vector gauge potential $A_\mu=(\sqrt{3}e/r^2,0,0,0)$.
 The coupling constant $h$ given in
the metric solution (2.9)
  represents hairy charge and is given by
 \begin{equation}
h=\frac{64\pi G}{5}b_1n^{9},~~~n=\varepsilon \left(
-\frac{18}{5}\frac{b_{1}}{b_{0}}\right)^{1/6},
 \end{equation}
with $\varepsilon=-1,0,+1$ and extra condition
$10b_{0}b_{2}=9b_{1}^{2}$ which guarantees the metric solution
(2.9) to have a black hole topology. Also the scalar field
configuration of this charge takes the form of
$\phi(r)=\dfrac{n}{r^{1/3}}$. Thermodynamic quantities of this
hairy black hole are given  as follows [44].
\begin{equation}
Q=-\frac{\sqrt{3}\pi}{8}q
\end{equation}
\begin{equation}
M=\dfrac{3\pi}{8}m=\dfrac{3\pi}{8}\left(  r_{+}^{2}-\dfrac{h}{r_{+}}+\frac{q^2}{r_{+}^{2}}%
+\dfrac{r_{+}^{4}}{\ell^{2}}\right)  \,, \label{M2}%
\end{equation}
\begin{equation}
S=\int^{r_+}_0\frac{1}{T}\left(\frac{\partial M}{\partial
r_+}\right)dr_+=2\pi^2(\frac{r_+^3}{4}-\frac{5}{8}h),
\end{equation}
\begin{equation}
T(r_{+})=\dfrac{1}{\pi\ell^{2}r_{+}^{4}}\left(-\frac{q^2 \ell^{2}}{2r_{+}}+\frac{h \ell^{2}}{4}%
+\frac{\ell^{2}}{2}r_{+}^{3}+r_{+}^{5}\right),
\end{equation}
where  the black hole event horizon $r_+$ is determined by solving
the horizon equation $f(r_{+})=0.$ The entropy computed in (2.13)
differs from the standard Bekenstein-Hawking formula.
 In fact it accords to the
Iyer-Wald's method [31, 33].  Since in the extended thermodynamics
the black hole mass behaves as the enthalpy, so by regarding
(2.11), (2.12), (2.13) and (2.14), we can reach to the first law
of extended thermodynamics  as follows.
\begin{equation}
dM=TdS+VdP+Kdh+\Phi dQ,
\end{equation}%
where the thermodynamic volume $V$ is conjugate potential of the
pressure in 5-dimensional space time given by $P=3/4\pi\ell^2$ .
One can compute $V$ and other conjugate potentials for hairy and
electric charges indicated by $K$ and $\Phi$, respectively as
follows:
\begin{eqnarray} \ V&=&\left(\frac{\partial M}{\partial
P}\right)_{r+}=\frac{\pi^2}{2}~r_+^4,
\\
\ K&=&\bigg(\frac{\partial M}{\partial
h}\bigg)_{r+}=\frac{\pi}{16\ell^2r_+^5}\big(20r_+^6+4r_+^4\ell^2+5\ell^2hr_+-10q^2\ell^2\big),
\\
\ \Phi&=&\left(\frac{\partial M}{\partial
Q}\right)_{r+}=-\frac{2\sqrt{3}}{r_{+}^2}q.
\end{eqnarray}
The generalized Smarr relation regarding the scaling argument
reads
\begin{equation}
2M=3TS-2VP+3hK+2\Phi Q.
\end{equation}
Now we can obtain the equation of state for hairy black hole by
using Hawking temperature (2.14) and definition of specific
volume, $v=\frac{4}{3}r_{+}$ [34,55] such that,
\begin{equation}
P=\frac{T}{v}-\frac{2}{3\pi v^2}-\frac{64}{81}\frac{h}{\pi
v^5}+\frac{512}{243}\frac{q^2}{\pi v^6}.
\end{equation}
The critical points are obtained by solving the equations
\begin{equation}
  \left.\frac{\partial P}{\partial v}\right|_{T=T_{cr}}=\left.\frac{\partial ^2P}{\partial v^2}\right|_{T=T_{cr}}=0,
\end{equation} which in absence of the Maxwell field $q=0$ they reduce to the solutions
\begin{equation}
v_{cr}=\frac{4}{3}(-5h)^\frac{1}{3},
\end{equation}
\begin{equation}
T_{cr}=-\frac{3}{20}\frac{(-5h)^\frac{2}{3}}{\pi h},
\end{equation}
\begin{equation}
P_{cr}=\frac{9}{200\pi}(\frac{-\sqrt{5}}{h})^\frac{2}{3}
\end{equation}
 for which  we must be choose $h<0$ to have critical points with real values. In general for $q\neq0$
 one can obtain numerical solutions for (2.21) [32]. Regarding these results the situations are divided into two branches:
 If $h<0$  then  the critical points would be physical for all values of $q$,
 but if $h>0$ then the equation of state admits a single inflection point.
  To have physical critical points the corresponding entropy should not to be negative for this inflection point.
 With numerical analysis,  one can infer that all critical points defined on
 $h\approx1.3375q^{3/2}$ are physical points with zero entropy.
 In the other side, all critical points under this curve in $(h,q)$ parameter space are physical with positive entropy
  (see figure 3 in [32]).

\section{Holography heat engines}
In this section we study hairy black holes in $AdS_5$ spacetime
 which can be considered to be as
the working substance of the heat engine. For this purpose we
consider a rectangular cycle in $P-V$ diagram as shown in figure
1. As we explained in the introduction, variation of cosmological
parameter (namely the pressure of the AdS spacetime) is viewed as
the change of size of the AdS spacetime. All these changes cause
to behave the curved spacetime as a thermodynamical system in
which $P-V$ diagram is given in the figure 1. The cycle is made up
of four processes: two constant volume process es (isocore) and
two constant pressure process es (isobar). During this rectangular
cycle, the system receives heat during $b\rightarrow c$ and
$c\rightarrow d$, so the net amount of heat
 transferred to the system is $Q_{H}=Q_{b \to c}+Q_{c\to d}$. Then the system expels heat during $d \to a$  and $a \to b$.  Because the process
  denoted by
 $b \to c$
 takes place in constant volume, the amount of heat entered to the system would be equal to the changing in the internal
 energy   of the system given by
 $\Delta U=Q_{bc}=C_{v} \Delta T_{b \to c}$. But since the volume and entropy of a black hole are both depend only on the radius of the horizon,
 we can conclude that the specific heat at constant volume vanishes so the system does not receive any heat during the $b \to c$ process.
  On the other side, the heat of the system absorbed during the isobaric process
 $c \to d$ can also be obtained from $Q_{cd}=C_{P} \Delta T_{b \to c}$.
  So the total heat entered into the system in this cycle is given by the following relation.
 \begin{equation}
Q_H=\int^{T_d}_{T_c}C_PdT,
\end{equation}
\begin{figure}[tbp] \centering
    \includegraphics[width=8cm,height=7cm]{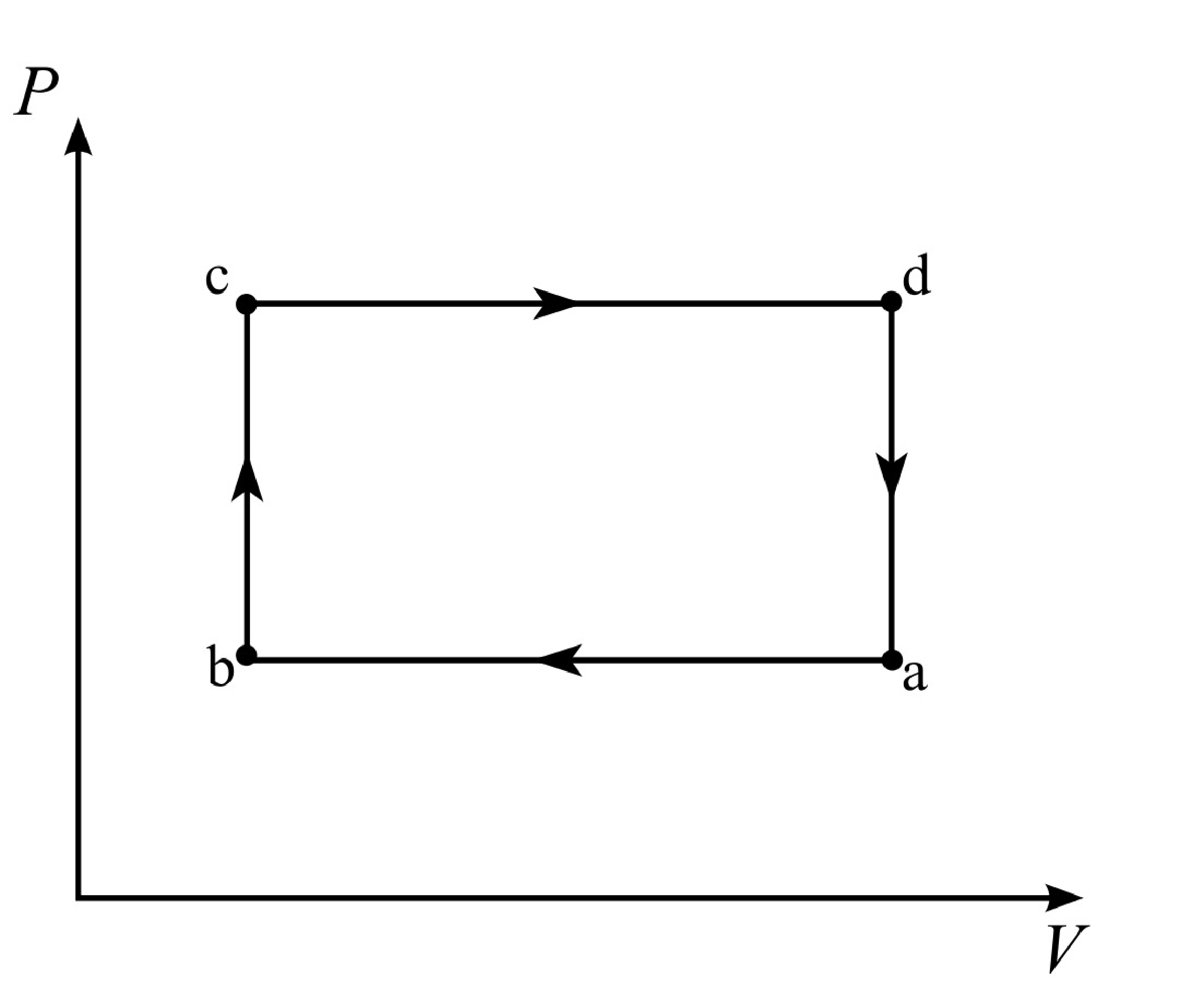}
\caption{\label{fig:1}The heat engine cycle in $P-V$
 plan for a thermodynamical system
}
\end{figure}
where  the specific heat at constant pressure $C_P$ is defined by
\begin{equation}
C_P=T\left(\frac{\partial S}{\partial T}\right)_{P}.
\end{equation}
 Applying metric potential (2.9)
and horizon equation $f(r_+)=0$ and thermodynamic quantities
(2.11), (2.12), (2.13) and (2.14) one can show that the above heat
capacity leads to the following form.
\begin{equation}
C_P=\frac{3(2PA^2 \pi^2+12h\pi^4 B^{\frac{1}{3}}+3(A\pi)^\frac{7}{3}-48q^2 \pi^5)}{16(240q^2 \pi^5+2PA^2\pi^2-48h\pi^4 B^\frac{1}{3}-3\pi
 (A\pi)^{\frac{4}{3}})},
\end{equation}
where $A=20\pi^2 h+16S$ and $B=20\pi^3 h+16S$. The inflow heat
relation will be simplified by using (3.1) and (3.2) in the
expansion process $c \to d$ if we hold all other quantities as
constants.  In the latter case we will have
\begin{equation}
Q_H=\int^{S_d}_{S_c}C_P \left(\frac{\partial T}{\partial S}\right)_PdS=\int^{S_d}_{S_c}TdS=\int^{M_d}_{M_c}dM=M_d-M_c.
\end{equation}%
 For hairy $AdS_5$ black holes,
from (2.12) we have
\begin{equation}
M_i=\frac{3\pi}{8}\big(r_{+_{i}}^2-\frac{h}{r_{+_{i}}}+\frac{q^2}{r_{+_{i}}^2}+\frac{4\pi}{3}Pr_{+_{i}}^4\big),
\end{equation}
  where the event horizon radiuses
$r_{+_{i}}=\frac{1}{2\pi^{2/3}}A_i^\frac{1}{3}$ with $i=c,d$ and
$A_{i}=20\pi^2 h+16S_i$  could be obtained from (2.13). So the net
amount of heat which the system receives is
\begin{eqnarray}
\nonumber Q_{H}=M_{d}-M_{c}=\frac{3\pi}{8}\bigg(\frac{1}{4\pi^{4/3}}(A^\frac{2}{3}_d-A^\frac{2}{3}_c)-2\pi^{2/3} h(A^\frac{-1}{3}_d-A^\frac{-1}{3}_c)\\
+4\pi^{4/3} q^2 (A^\frac{-2}{3}_d-A^\frac{-2}{3}_c)+\frac{P}{12\pi^{5/3}}(A^\frac{4}{3}_d-A^\frac{4}{3}_c)\bigg).
\end{eqnarray}
 On the other side, the work
performed by engine equals to the area enclosed by the cycle which
is computed regarding figure 1 as follows:
\begin{equation}
W=(V_d-V_c)(P_c-P_b),
\end{equation}%
where subscripts denote to the points shown in the figure 1. So by
attention to (2.16) the equation (3.7) reads
\begin{equation}
W=\frac{(A_{d}\pi)^\frac{4}{3}-(A_{c}\pi)^\frac{4}{3}}{32\pi^2}(P_{c}-P_{b}).
\end{equation}%
Finally we can demonstrate the performance of the heat engine by a
thermal efficiency $\eta$  as follows.
\begin{equation}
\eta=\frac{W}{Q_H}.
\end{equation}%
To investigate the effect s of hairy and Maxwell charges on the
performance of heat engine it would be useful to set one charge
with unit and let the other one
 changes in  the
ranges with positive Hawking temperature. In figures $2.a$ and
$2.b$ we show the maximum value of charge with a red spot after
which Hawking temperature will be negative.
As we can see from this maximum charge the thermal efficiency has also its maximum value.\\
In figure $2.a$ we plotted the heat engine efficiency
 versus the  Maxwell charge by
  holding $P_c=2$, $P_b=1$ , $S_c=8$, $S_d=10$ for
various fixed hairy charge. The diagram shows that the efficiency
grows by increasing the Maxwell charge for any fixed hairy black
holes until  it ceased at a maximum value of point
$(q_{max},\eta_{max})$ after which temperature would be
negative.\\
 Regarding numerical analysis   which  we pointed it out at end of the section 2 the
 critical points might be un-physical
  for  $q\neq0$
 where $h>0$ and the entropy
reaches to some negative values. For example the blue line in
figure $2.a$ is fixed at $h=1$,
 so all physical critical points must be restricted by $q_{phys}\geqslant0.6464862821$.
  The green line corresponds to AdS-RN black hole solution which is independent
 of the hairy charge [12].
 In the latter case all points from
$q=0$ until a maximum value are physical. As we can see the
existence of hairy charge increases the maximum point
$(q_{max},\eta_{max})$.  In other words the physical critical
points admit synchronously bigger values of electric charges and
efficiencies by increasing the hairy charge.
  On the other side, hairy charge causes a new restriction due to negative entropy
  and therefore physical critical points start from a
  point with $q_{phys}>0$.
\begin{figure}[h]
\centering
\subfigure[{}]{\label{1}
\includegraphics[width=.45\textwidth]{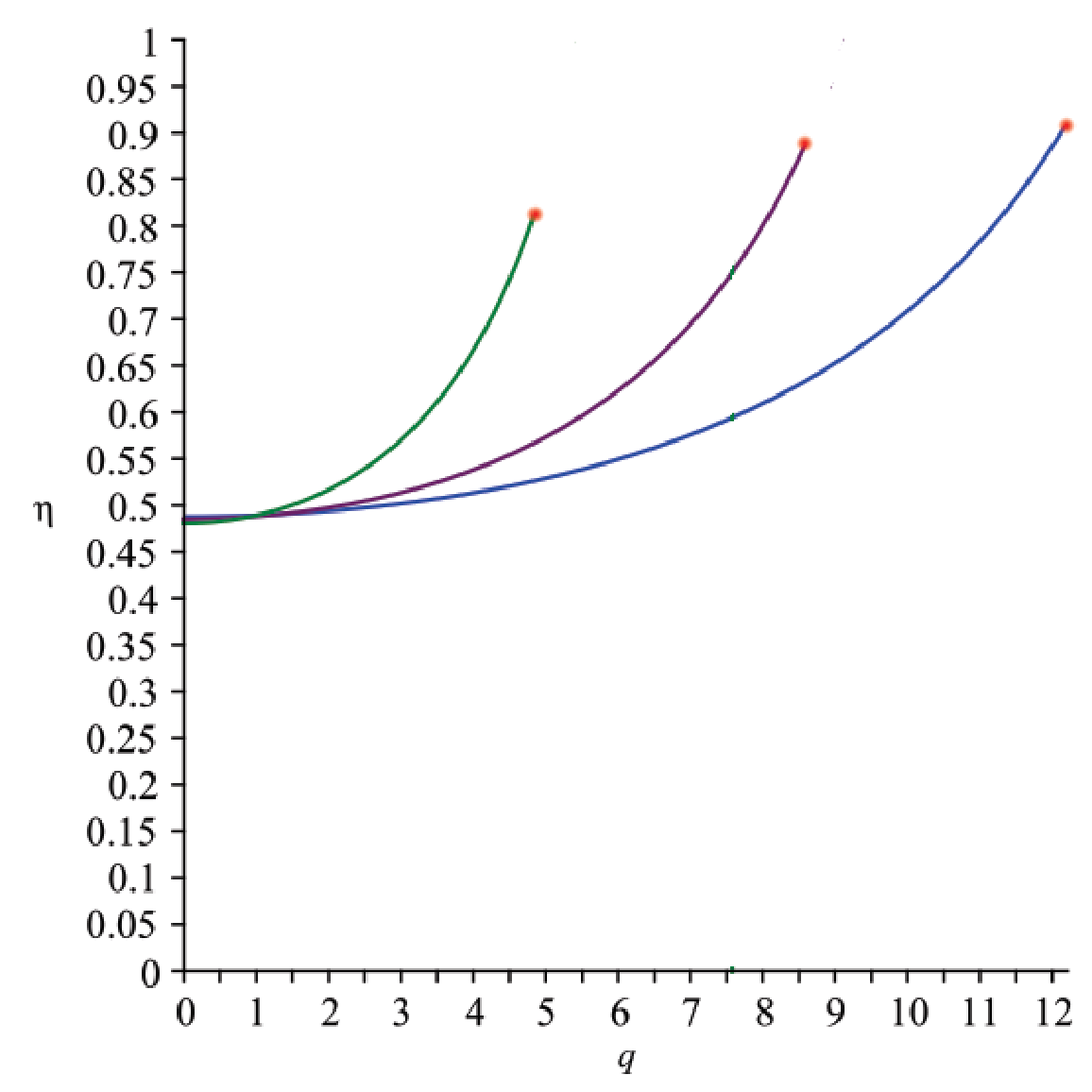}}
\hspace{1mm}
\subfigure[{}]{\label{1}
\includegraphics[width=.45\textwidth]{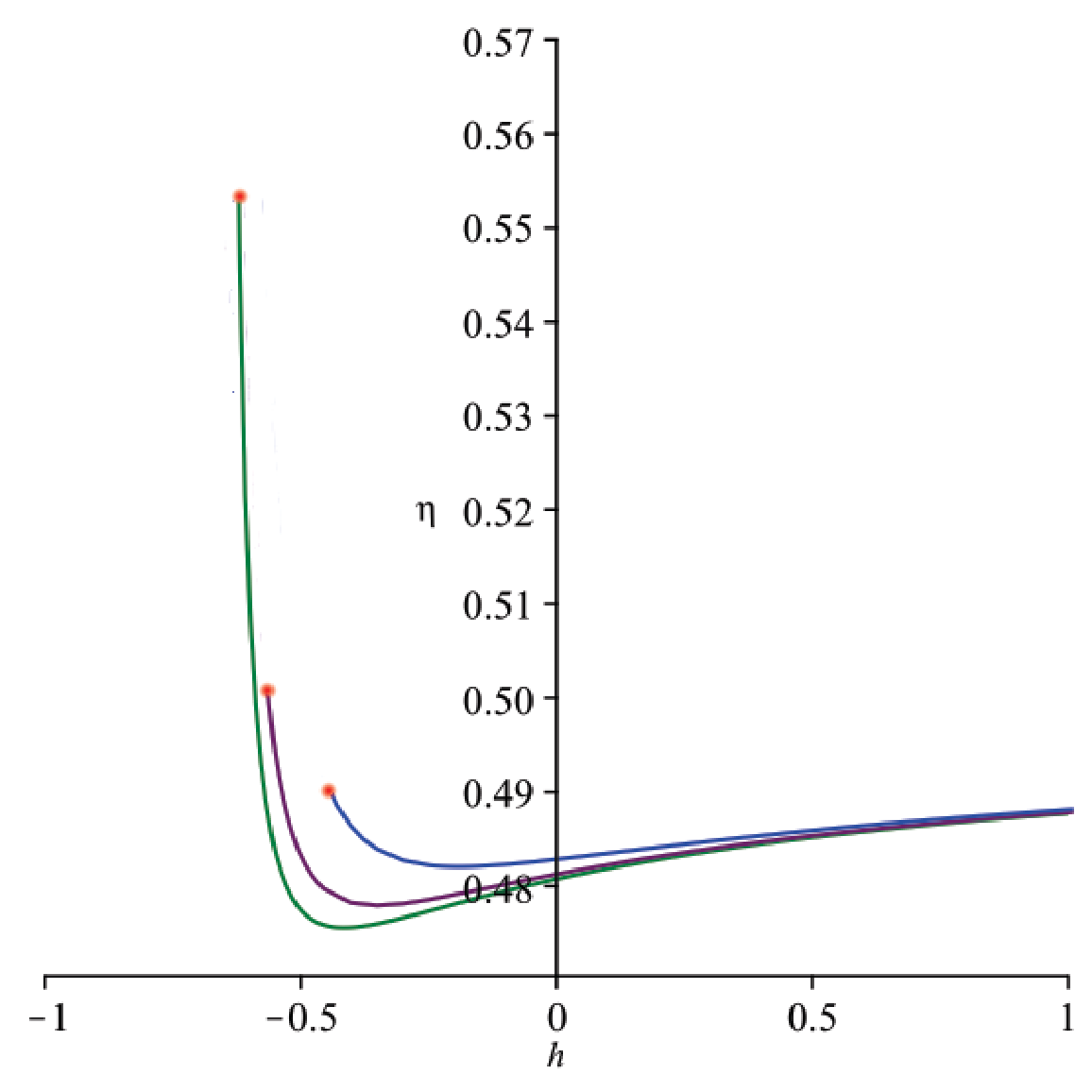}}
\caption{ Efficiency is plotted versus the Maxwell electric charge
$q$ in (a) and the hairy charge $h$ in (b) by fixing $h$ in (a)
and $q$ in (b) respectively. In $(a)$ diagrams reach to a maximum
point at $q>0$ while in (b) they reach to a maximum  point at
$h<0$. In (a) Green, purple and blue lines indicate fixed values
$h=0$, $0.5$ and $1$ respectively for which maximum point
 is $q_{max}=4.890,\eta_{max}=0.824$ for green line,  $q_{max}=8.572,\eta_{max}=0.885$ for
  purple line and  $q_{max}=12.234,\eta_{max}=0.915$ for blue line respectively.  In $(b)$
  Green, purple and blue lines indicate fixed values $q=0$, $0.25$ and $0.5$,
  respectively.}
\label{l}
\end{figure}
In figure $2.b$ we fix electric charge and let hairy charge
  to be change by holding $P_c=2$,
$P_b=1$ , $S_c=8$, $S_d=10$. As we can see there is a minimum
point at which system has a minimum efficiency and so in this case
efficiency behaves more complicated rather than previous case.
 The green line indicates case of
electric charge independent  at which (as we discussed before)
hairy charge must be have  negative value to have physical
critical points, so all positive values of $h$ are not physical
for this line.  Considering this condition the absolute value of
maximum hairy charge ($h_{max}=-0.6066$ with $\eta_{max}=0.553$)
and minimum point (with $h_{min}=-0.418$ and $\eta_{min}=0.476$)
both are physical. However for $q\neq0$, hairy charge could be
positive or negative, thus in contrary to  the case where $q=0$,
the physical critical points have $h>0$ provided that they satisfy
positive entropy condition. For instance for $q=0.5$ indicated
 by blue line, all critical points with $h>0.8425722021$ are not physical.
  This diagram shows that by increasing electric charge,
 both maximum and minimum points are satisfied for smaller hairy charges
  but with different behavior in their efficiencies.
  Actually, the maximum point will has smaller efficiency by increasing electric charge while the
  efficiency grows for the minimum point. Indeed, for fixed large electric charge, their efficiencies reach together. \\
 We can compare
the heat engine efficiency $\eta$ with the Carnot efficiency
$\eta_{c}=1-\frac{T_{low}}{T_{high}}$ where $T_{low}$ and
 $T_{high}$ are the lowest and the highest temperature of the heat engine in cycle under consideration.
  Diagrams $3.a$ and $3.b$ show that when we holed electric (hairy) charge with a constant
 value, then $\eta _{h}\rightarrow 1$
 at maximum value of hairy (electric) charge.
 Diagram  $3.a$ shows that for
fixed larger $q$ we have smaller absolute value for $h_{max}$ for
which $\eta_{c}\rightarrow1$, in contrary, $3.b$ shows that by
increasing the hairy charge $h$ we will have bigger $q_{max}$.
There is similar analysis for relations between $q$ and $h$ with
physical critical points as we discussed already. The ratio
$\eta/\eta_{c}$ is plotted versus the hairy and Maxwell electric
charges in figures $4.a$ and $4.b$, respectively. In
  figure $4.a$ with $h=1$ one can see decreasing of this ratio by increasing the Maxwell charge, while with $q=1$ in figure $4.b$ we see inverse
  behavior
  for it. One can see that
   $\eta/\eta_{c}$ increases by raising the hairy charge.
   For large h it approaches to 1 which means that $\eta\rightarrow\eta _{c}$.\\
\begin{figure}[h]
\centering
\subfigure[{}]{\label{1}
\includegraphics[width=.45\textwidth]{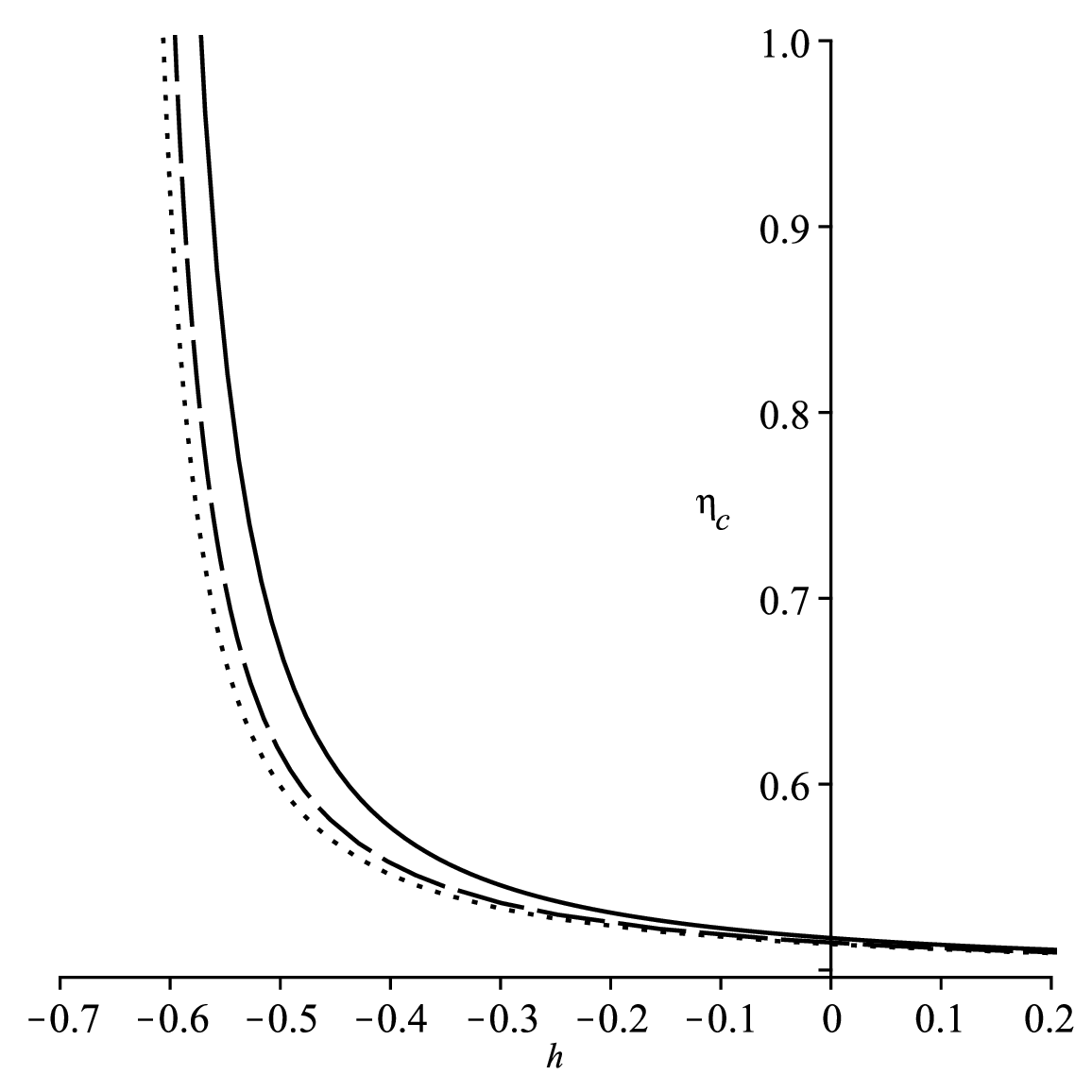}}
\hspace{1mm}
\subfigure[{}]{\label{1}
\includegraphics[width=.45\textwidth]{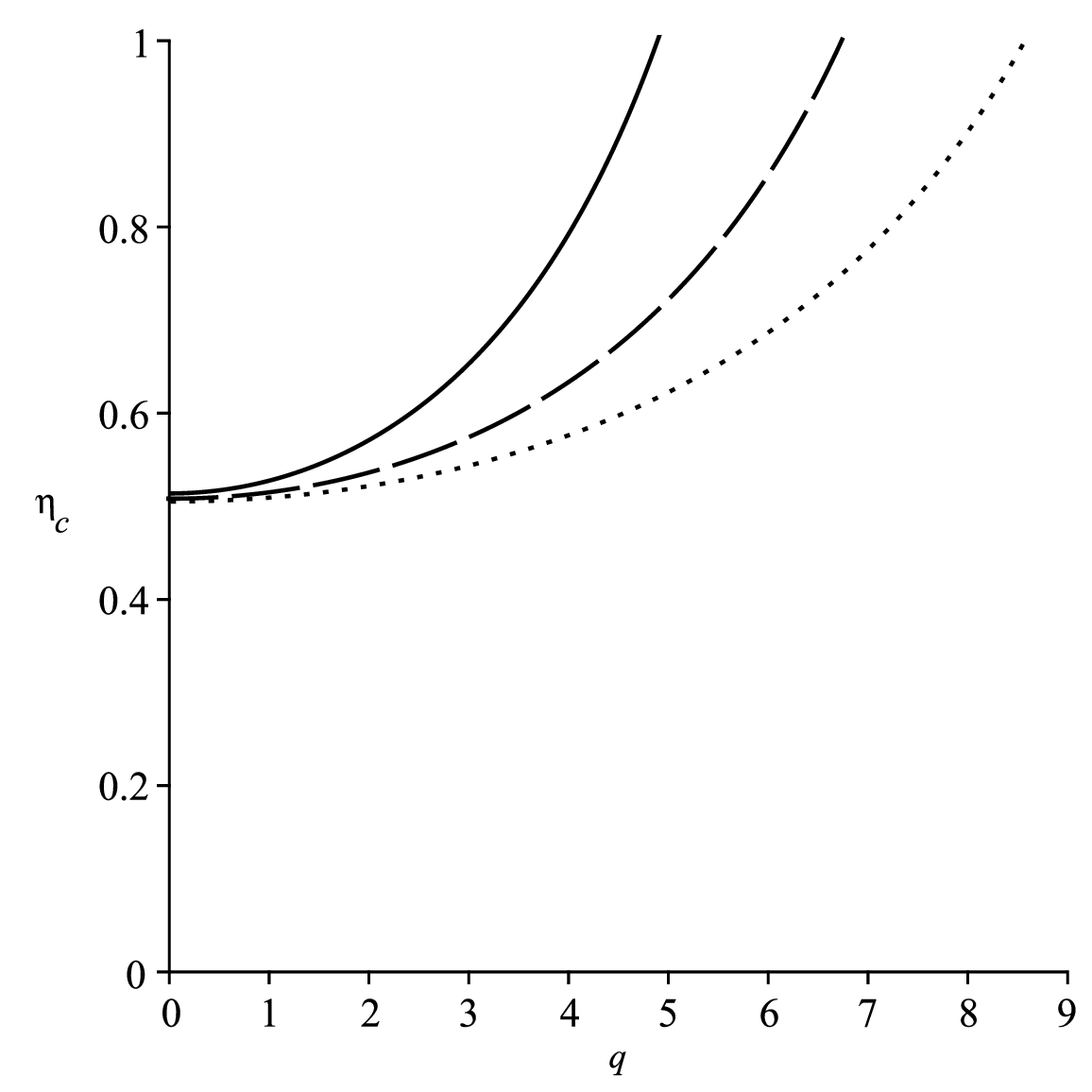}}
\caption{ Behavior of the Carnot efficiency is plotted versus the
charges for fixed $q$ in $(a)$ and fixed $h$ in $(b)$
respectively. Solid, dash and dotted lines represent
 $q=0.5, 0.25, 0$ in $(a)$, and $h=0, 0.25, 0.5$ in
$(b)$ respectively.   } \label{l}
\end{figure}
\begin{figure}[h]
\centering
\subfigure[{}]{\label{1}
\includegraphics[width=.45\textwidth]{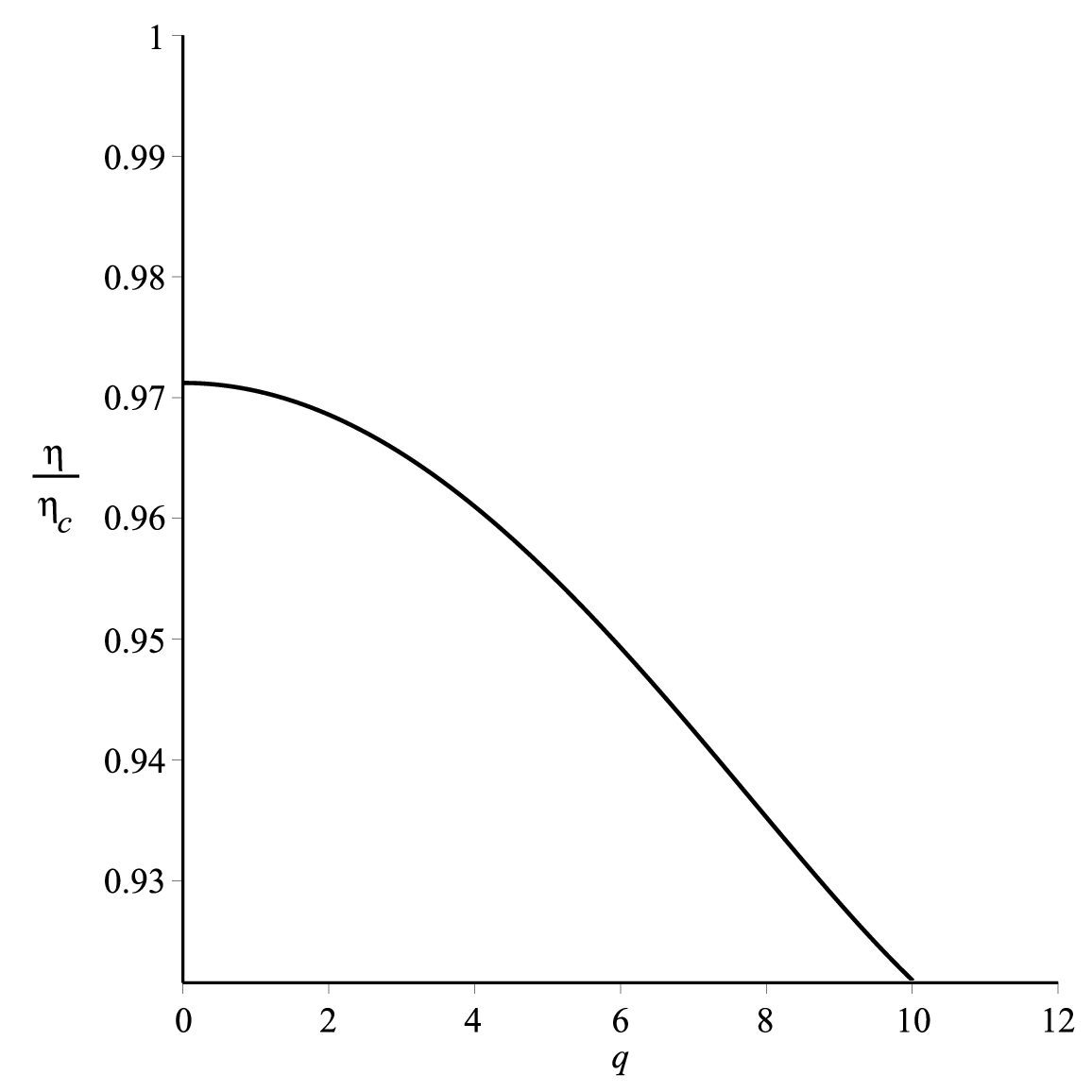}}
\hspace{1mm}
\subfigure[{}]{\label{1}
\includegraphics[width=.45\textwidth]{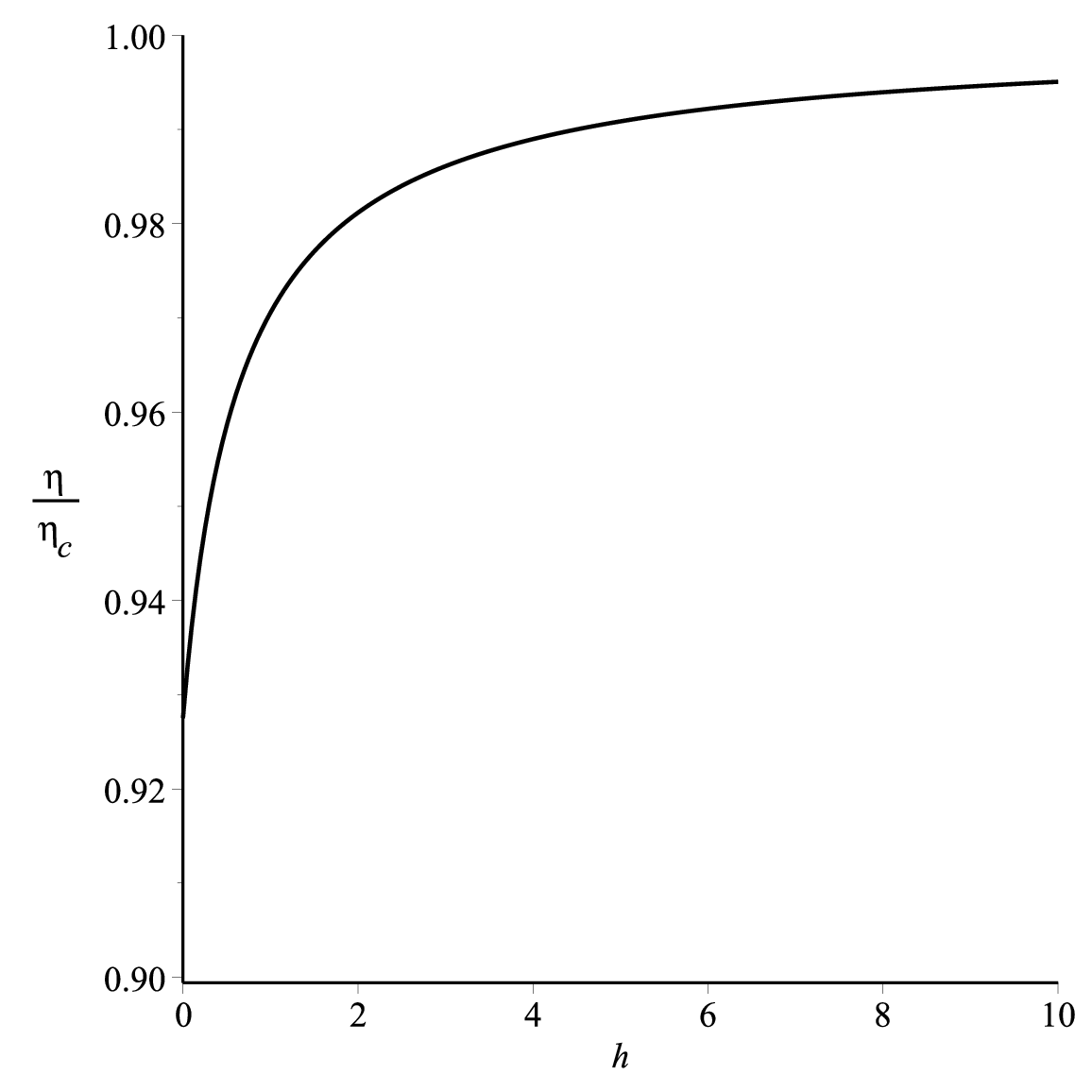}}
\caption{ Diagram of ratio $\eta/\eta_c$ is plotted versus the
charges for $P_c=2$, $P_b=1$, $S_c=8$ and $S_d=10$ where $h=1$ in
(a) and $q=1$ in (b).} \label{l}
\end{figure}
$\bullet$\emph{Approaching the Carnot Limit:} As it studied in
[56,57] when we put one of the corner of rectangular cycle in
 the critical points of the system or near to
them, the efficiency of heat engine could be approached to the
Carnot efficiency having the finite power. In [14] we can see good
results for charged-AdS black hole in the limit of large charge.
Following this work we put the critical point $(P_{\rm cr}, V_{\rm
cr})$ in corner $a$ and
 choose the boundary conditions as
\begin{eqnarray}
 \nonumber  P_a=P_b =P_{\rm cr},~~~ P_c=P_d  =  \frac{3}{2}P_{\rm cr}, \\
  V_d = V_a=V_{\rm cr},~~~  V_b = V_c= V_{\rm cr}\big(1-\frac{L}{h^{2/3}}\big),
\end{eqnarray}
 in which the constant parameter
$L$ comes from dimensional analysis. For large values of the
charge we can put $\alpha\equiv\frac{L}{h^{2/3}}\rightarrow0$. The
work done by the engine could be simply calculated by the area of
$\Delta V\Delta P$  diagram in the cycle under consideration
  as follows.
\begin{equation}
 W =  \frac{1}{2} P_{\rm cr}V_{\rm cr}\alpha=\frac{9\times5^{2/3}}{160}\pi L.
\end{equation}
Note that the work done in cycle under the above considerations is
finite and independent of hairy charge. The heat
 absorbed by the system is given by
\begin{equation}
 Q_{H} =  M_{2}-M_{1}=\frac{3\pi}{8}(5H)^{2/3}\bigg((1-\alpha)^{1/2}+\frac{1}{20}(9\alpha+24-4(1-\alpha)^{-1/4})\bigg),
\end{equation}
where by attention to the negativity of hairy charge
 for black hole system free of
Maxwell electric charge we put $H=-h$ as the absolute value of
$h$. For large values of the hairy charge for which $\alpha\to 0$,
the absorbed heat energy by the system takes the following form
\begin{equation}
Q_{H}=\frac{27\pi}{80}(5H)^{2/3}\alpha+\frac{9\pi}{256}(5H)^{2/3}\alpha^2+\frac{15\pi}{1024}(5H)^{2/3}\alpha^3+\mathcal{O} \left({\alpha}^{4} \right).
\end{equation}
Therefore, the efficiency of the heat engine (3.9) leads to the
following form
\begin{equation}
\eta=\frac{1}{6}\bigg(1-\frac{5}{48H^{2/3}}-\frac{25}{768H^{4/3}}+\frac{875}{110592H^2}+...\bigg)
\end{equation}
 where we used (3.11) and put
$L=1.$  In the other side, the Carnot efficiency as we discussed
before depends just on the highest $(T_H)$ and the lowest $(T_C)$
temperatures of the heat engine where they happen at the corners
$d$ and $b$, respectively. In limits of the large hairy charge we
obtain
\begin{equation}
T_H=\frac{9}{50\pi}\big(\frac{25}{H}\big)^{1/3},
\end{equation}and \begin{equation}
T_C=\frac{3}{20\pi}\big(\frac{25}{H}\big)^{1/3}-\frac{1}{640\pi}\big(\frac{25}{H^7}\big)^{1/3}+\cdots
.
\end{equation}
 Substituting (3.15) and (3.16)
into the definition of the Carnot efficiency one can infer
\begin{equation}
\eta_{c}=1-\frac{T_C}{T_H}=\frac{1}{6}\bigg(1+\frac{5}{96 H^2}+...\bigg).
\end{equation}
 By attention to (3.14) and (3.17)
we can conclude $\eta\approx\eta_c$ for large hairy charge $H\to
\infty$ (see figure 5). \\
By attention to  the critical pressure (2.24) we can obtain $P\sim
H^{-2/3}$ and so $\tau\sim H^{2/3}$.  According to the work [14]
one can infer that $\tau$ is the necessary time to complete a
cycle scales at a finite hairy charge leading to a finite power
$W/\tau$ in which $W$ is work given by (3.11) (see [57,58]) such
that
\begin{equation}
\frac{W}{\tau}\leq\bar{\Theta}\frac{\eta(\eta_c-\eta)}{T_C}.
\end{equation}
 In the above equation
$\bar{\Theta}$ is a model dependent constant and the right hand
side efficiency function would be expanded
 versus the large hairy charge
$H(\equiv-h)$ as follows:
\begin{equation}
\frac{\eta(\eta_c-\eta)}{T_C}=\frac{5^{4/3}\pi}{1296}\bigg(H^{-1/3}+\frac{5}{24}H^{-1}+\frac{827}{2304}H^{-5/3}+...\bigg).
\end{equation}
\begin{figure}[tbp] \centering
    \includegraphics[width=8cm,height=7cm]{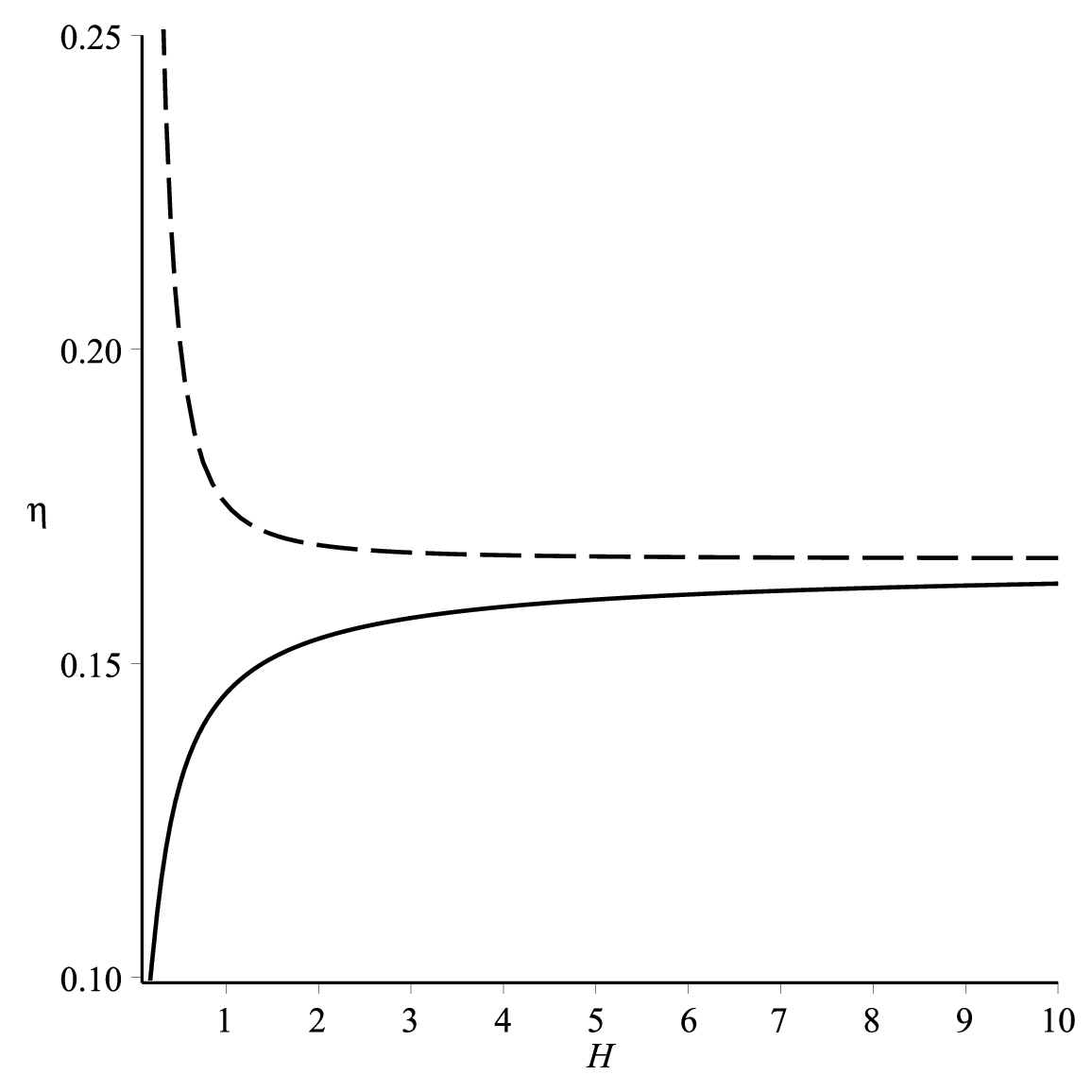}
\caption{ Diagrams of the heat engine efficiency (solid line) and
the Carnot efficiency (dash line) are plotted versus the large
hairy charge $H$ for a black hole solution with zero Maxwell
electric charge $q=0. $ }
\end{figure}
When  the hairy charge goes to infinity and
 so $\eta=\eta_c$
 then the power (3.18) vanishes as
we expect. The universal trade-off relation (3.18) (see [57]) as a
bound between power and efficiency would be satisfied for any
values of $h$.
\section{Conclusions}
 As thermodynamics of the black
holes with the scalar hair configurations are favored more than of
the non-hairy ones, thus we studied thermodynamics of black holes
in $AdS$ space time which have both electric charge and hairy
charge. In the latter case the cosmological constant plays as
thermodynamic variable.  As the thermodynamic processes of black
hole system construct a closed cycle then it operates similar to a
heat engine. So we can study this black hole heat engine as an
implement of the black hole thermodynamics.  We study
thermodynamics of the hairy black holes in $AdS_5$ space-time due
to the interesting features of its phase transition. From [53] we
know  that the Maxwell electric and hairy charges are bounded
together and the sign of entropy depends on this bound.
 Looking to the work  [53] we
understand that positive critical points are divided into physical
and un-physical branches with which represent positive and
negative entropies, respectively.
 We found out from the discussion
leading to the figure.2 that for a fixed hairy charge the
efficiency of black hole solution (which acts as a heat engine)
changes by the Maxwell electric charge. Maximum value of the
efficiency is happened at a maximum charge after which we have
unacceptable negative temperature. This maximum point moves to
larger values by choosing bigger fixed hairy charges.
   It could be seen from
figure ($2.a$) the blue line has the biggest value of fixed $h$
and so it has biggest maximum points for the Maxwell charge and
efficiency. Indeed for bigger Maxwell charges in these diagrams
the system enters to an un-physical situation which is not
acceptable.  In the other side, we found  also a minimum value of
 the efficiency in the case of fixed electric
charge  which could be seen in figure ($2.b$) graphically.
 We concluded that if we choose
$q=0$ then all critical points for $h>0$ will be un-physical,
however for all other fixed $q\neq0$ we have physical critical
points with positive hairy charges for which the entropy remains
positive.  Maximum and minimum critical values are physical for
all values of fixed electric charges and they have different
behavior by increasing the Maxwell electric charge. By increasing
$q$, maximum (minimum) point happens at smaller hairy charges but
with smaller(larger) efficiency.
 It means for example if we choose
larger values for a fixed Maxwell electric charge then the hairy
charge
for both maximum and minimum point decreases, but we can see this change make the value of efficiency smaller(larger) for the maximum(minimum) point.\\
We also compared this efficiency with Carnot efficiency
 and conclude that  the ratio  $\eta/\eta_{c}$  will
  decrease by changing of Maxwell charge when $h=constant$.  On the other side when we choose a fixed hairy
  charge it could be seen that $\eta\rightarrow\eta_c$ for large $h$.
  Regarding [14] and by putting one corner of rectangular cycle of heat engine process on or near critical points of the system the efficiency of heat
   engine reaches to the Carnot efficiency having a finite power. We studied it in large hairy charge limit when the power vanishes.\\
  Previously,  thermodynamic processes were studied for various black hole systems as heat engines, but by considering a conformal scalar hair arising from coupling of a real scalar field with dimensionally extended Euler densities, can help us to explore more profound in thermodynamic behavior of black holes.
  Considering the hairy charge could be also important when we try to found out its behavior in field theory side in AdS/CFT proposal.  It could be correspond to some kind of hyper charges or baryon numbers.
   As the black hole has scalar hair it could be make contact with something more akin to a superconducting phase like which we can explore in [59] and [60].
\section{Acknowledgment}
  We would like  to thank the editor Prof. \textit{Stephan Stieberger} and anonymous reviewers for their valuable guidance and also  $Pavan~Kumar~Yerra$
 from \textit{Indian Institute of Technology Bhubaneswar} for our discussions and his useful comments
 which caused to improve the work. \\

  \vskip .5cm

 \noindent
  {\bf References}\\
\begin{description}
\item[1.] J. D. Bekenstein,  "Black holes and entropy." Phys. Rev. D7, 2333, (1973).
\item[2.] S. W. Hawking,  "Particle creation by black holes." Commun. Math. Phys. 43, 199, (1975).
\item[3.] S. W. Hawking and P. Don "Thermodynamics of black holes in anti-de Sitter space", Commun. Math. Phys. 87, 577,
(1983).
\item[4.] A. Chamblin,  R. Emparan, C. V. Johnson and R. C. Myers "Charged AdS black holes and catastrophic holography." Phys. Rev.
D60, 064018, (1999).
\item[5.] A. Chamblin, R. Emparan, C. V. Johnson and R. C. Myers. "Holography, thermodynamics, and fluctuations of charged
AdS black holes." Phys. Rev. D60, 104026, (1999).
 \item[6.] L. Smarr, "Mass  Formula
for  Kerr  BlackHoles",Phys. Rev. Lett 3, 71 (1972), Erratum
ibid.,30, 521 (1973).
 \item[7.] S.Q. Hu, X.M. Kuang and Y.C. Ong, "A Note on Smarr Relation and Coupling
Constants" ,Gen. Relativ. Gravit. 51, 55 (2019), gr-qc/1810.06073
\item[8.] M. M. Caldarelli, G. Cognola, and D. Klemm. "Thermodynamics of Kerr-Newman-AdS black holes and conformal field
theories." Class. Quant. Grav. 17, 399, (2000).
\item[9.] D. Kastor, S. Ray and J. Traschen "Enthalpy and the mechanics of AdS black holes." Class. Quant. Grav. 26, 195011, (2009).
\item[10.] L. Girardello, M. Petrini, M. Porrati, and A. Zaffaroni.
"Novel local CFT and exact results on perturbations of N=4 super
Yang Mills from AdS dynamics." JHEP 1998, 022, (1999).
\item[11.] J. Distler and F. Zamora. "Non-supersymmetric conformal field theories from stable anti-de Sitter spaces",
Adv. Theor. Math.Phys.2, 1405,(1999), hep-th/9810206.
\item[12.] C. V. Johnson "Holographic heat engines." Class. Quant. Grav. 31, 205002, (2014).
\item[13.] C. V. Johnson,  "An exact efficiency formula for holographic heat engines" Entropy 18, 120, (2016).
    \item[14.] C. V. Johnson and F. Rosso. "Holographic heat engines, entanglement entropy, and renormalization group flow." Class.Quant.Grav. 36, 015019
     (2019), hep-th/1806.05170.
\item[15.] C. V. Johnson "An Exact model of the power-to-efficiency trade-off while approaching the Carnot limit." Phys. Rev. D 98, 026008, (2018).
\item[16.] C. V. Johnson, "Gauss-Bonnet black holes and holographic heat engines beyond large N." Class. Quant. Grav. 33, 215009, (2016).
\item[17.] C. V. Johnson,  "Born–Infeld AdS black holes as heat engines." Class. Quant. Grav. 33, 135001, (2016).
\item[18.] A. Belhaj, M. Chabab, H. El Moumni, K. Masmar, M. B. Sedra and A. Segui. "On heat properties of AdS black
 holes in higher dimensions." JHEP 2015, 149, (2015).
\item[19.] B. Chandrasekhar and P. K. Yerra. "Heat engines for dilatonic Born–Infeld black holes." Eur. Phys. J. C 77, 534, (2017).
\item[20.]  A. Chakraborty and C. V. Johnson, "Benchmarking Black Hole Heat Engines,â€, Int. J. Mod. Phys. D28, 1950012 (2019), hep-th/1612.09272.
\item[21.] A. Chakraborty and C. V. Johnson. "Benchmarking Black Hole Heat Engines, II",    Int. J. Mod. Phys. D28, 1950006 (2019),  hep-th/1709.00088.
\item[22.] R. A. Hennigar, F. McCarthy, A. Ballon and R. B. Mann. "Holographic heat engines: general
considerations and rotating black holes." Class. Quant. Grav. 34,
175005, (2017).
\item[23.] C. V. Johnson, "Taub-Bolt heat engines." Class. Quant. Grav. 35, 045001, (2018).
\item[24.] H. Liu, and X. H. Meng. "Effects of dark energy on the efficiency of charged AdS black holes as heat engines." Eur. Phys. J. C 77, 556, (2017).
\item[25.] J. X. Mo and G. Q. Li "Holographic heat engine within the framework of massive gravity." JHEP 122, (2018).
\item[26.] S. H. Hendi, B. E. Panah, S. Panahiyan, H. Liu, and X. H. Meng. "Black holes in massive gravity as heat engines." Phys. Lett. B 781,
40, (2018).
\item[27.] S. W. Wen, and Y. X. Liu. "Charged AdS black hole heat engines."  Nucl. Phys. B 946, 114700 (2019) gr-qc/1708.08176.
\item[28.] F. Rosso,  "Holographic heat engines and static black holes: a general efficiency formula", Int.J.Mod.Phys. D28 (2019),  hep-th/1801.07425.
\item[29.] J. X. Mo and S. Q. Lan. "Phase transition and heat engine efficiency of phantom AdS black holes" Eur. Phys. J. C.78, 666 (2018),
gr-qc/1803.02491.
\item[30.] N. Bocharova, K. Bronikov, and V. Melnikov, ``An exact solution of the system of Einstein equations and mass-free scalar field``, Vestn. Mosk. \textit{Univ. Fiz. Astronom}. 6, 706 (1970)

\item[31.] B. C. Xanthopoulos and T.E. Dialynas, "Einstein gravity coupled to a massless conformal scalar field in arbitrary space‐
time dimensions." J. Math. phys.33, 1463, (1992).
\item[32.] J. D. Bekenstein,  ``Black holes with scalar charge``,
Ann. Phys. (N.Y.) 91, 75, (1975).
\item[33.] R. G. Cai, L. M. Cao, L. Li, and R. Q. Yang. "PV criticality in the extended phase space of Gauss-Bonnet black holes in AdS space."
JHEP 2013, 5, (2013).
\item[34.] C. Klimcik, "Search for the conformal scalar hair at arbitrary D" J. Math. phys. 34, 1914, (1993).
\item[35.] C. Martinez and J. Zanelli, "Conformally dressed black hole in 2+ 1 dimensions." Phys. Rev. D 54, 3830, (1996).
\item[36.] C. Martinez, J. P. Staforelli and R. Troncoso, "Topological black holes dressed with a conformally coupled scalar
field and electric charge." Phys. Rev. D 74, 044028 (2006).
\item[37.] C. Martinez, R. Troncoso, and J. Zanelli. "de Sitter black hole with a conformally coupled scalar field in four dimensions." Phys. Rev. D 67,
 024008 (2003).
\item[38.] A. Anabalon and H. Maeda, "New charged black holes with conformal scalar hair." Phys. Rev. D 81, 041501 (2010).
\item[39.] S. Bhattacharya and H. Maeda. "Can a black hole with conformal scalar hair rotate?." Phys. Rev. D 89,  087501 (2014).
\item[40.] C. Martinez, ``\textit{In Quantum Mechanics of Fundamental Systems: The Quest for Beauty and Simplicity}``, edited by
M. Henneaux and J. Zanelli (Springer, New York, 2009).
\item[41.] F. Oliva, and S. Ray, "Conformal couplings of a scalar field to higher curvature terms." Class. Quant. Grav. 29, 205008 (2012).
\item[42.] G. Giribet, M. Leoni, J. Oliva and S. Ray, "Hairy black holes sourced by a conformally coupled scalar field in D dimensions." Phys. Rev. D 89,
085040 (2014).
\item[43.] M. Chernicoff, M. Galante, G. Giribet, A. Goya, M. Leoni, J. Oliva, and G. P. Nadal,
"Black hole thermodynamics, conformal couplings, and $R^2$ terms," JHEP 2016, 159, (2016).
\item[44.] G. Giribet,  A. Goya and J. Oliva. "Different phases of hairy black holes in AdS 5 space." Phys. Rev. D 91,  045031 (2015).
\item[45.] K. Bamba, "Thermodynamic properties of modified gravity theories." Int. J. Geomet. Meth. Mod. Phys. 13, 06 , 1630007, (2016);
 gr-qc/1604.02632.
\item[46.] S. Nojiri, and S. D. Odintsov, "Unified cosmic history in modified gravity: from $F(R)$ theory to Lorentz non-invariant models."
 Phys. Rep. 505, 59, (2011).
\item[47.] S. Capozziello and M. De Laurentis, "Extended theories of gravity." Phys. Rep. 509, 167, (2011).
\item[48.] S. Nojiri, S. D. Odintsov, and V. K. Oikonomou. "Modified gravity theories on a nutshell: inflation, bounce and late-time evolution."
Phys. Rep. 692 1, (2017).
\item[49.] S. Capozziello, and V. Faraoni. \textit{Beyond Einstein gravity: A Survey of gravitational theories for cosmology and astrophysics}.
 Vol. 170. Springer Science and Business Media, (2010).
\item[50.] K. Bamba,  and S. D. Odintsov. "Inflationary cosmology in modified gravity theories." Symmetry 7, 220, (2015); hep-th/1503.00442.
\item[51.] K. Bamba, S. Capozziello, S. Nojiri, and S. D. Odintsov,
 "Dark energy cosmology: the equivalent description via different theoretical models and cosmography tests." Astrophys. and Space Sci. 342, 155,
 (2012).
\item[52.] R. Hennigar and R. Mann. "Reentrant phase transitions and van der Waals behavior for hairy black holes." Entropy 17, 8056, (2015).
\item[53.] D. Kubiznak and R. B. Mann, P-V criticality of charged AdS black holes, JHEP 1207, 033, (2012).
 \item[54.] J. D. Bekenstein, ``Exact solutions of Einstein-conformal scalar equations``,  Ann. Phys. (N.Y.)82, 535 (1974)
\item[55.] M. Polettini,  G. Verley and M. Esposito "Efficiency statistics at all times: Carnot limit at finite power."
 Phys. Rev. Lett. 114, 050601, (2015).
\item[56.] M. Campisi and R. Fazio. "The power of a critical heat engine." Nat. Commun. 7, 11895, (2016).
\item[57.] N. Shiraishi, K. Saito and H. Tasaki, "Universal trade-off relation between power and efficiency for heat engines." Phys. Rev. Lett.
117, 190601, (2016).
\item[58.] N. Shiraishi and H. Tajima. "Efficiency versus speed in quantum heat engines: Rigorous constraint from Lieb-Robinson bound."
 Phys. Rev. E 96, 022138, (2017).
  \item[59.] S. A. Hartnoll, P. H.
Christopher and G. T. Horowitz, "Holographic superconductors."
JHEP 2008, 015, (2008), hep-th/0810.1563.
 \item[60.] S. S. Gubser, "Colorful Horizons with Charge in Anti-de Sitter Space." Phys. Rev. Lett. 101, 191601, (2008), hep-th/0803.3483.
\end{description}
\end{document}